# Approximating Clustering for Memory Management and request processing

[1]D.D.D.Suribabu, [2]Dr.T.Hitendra Sarma ,[3]Prof.B.Eswar Reddy

[1]Research Scholar, Department of CSE, JNTUA, Ananthapur, Andhra Pradesh..

[2]Professor, Department of CSE,Srinivasa Ramanujam institute of Engineering and Technology(SRIT), Ananthapur, Andhra Pradesh.

[3]Professor, Department of CSE, JNTUA,Kalikiri,Andhra Pradesh

**Abstract**- Clustering is a crucial tool for analyzing data in virtually every scientific and engineering discipline. There are more scalable solutions framed to enable time and space clustering for the future large-scale data analyses. As a result, hardware and software innovations that can significantly improve data efficiency and performance of the data clustering techniques are necessary to make the future large-scale data analysis practical. This paper proposes a novel mechanism for computing bit-serial medians. We propose a novel method, two-parameter terms that enables in computation within the data arrays

*Keywords*: Data clustering, bit-serial medians

## 1. Introduction

Data Clustering is one of the most fundamental compo-nents in learning and understanding the natural structure of data sets. Clustering techniques are increasingly used in science and engineering for important applications, such as precision medicine World Wide Web, machine learning, self-driving cars , business marketing , and economy . Due to recent advances in sensor and storage technologies, as well as the significant growth in the applications of unsupervised learning, data clustering has more terms to be considered.

K-means is one of the most commonly used algorithms for solving data clustering problems in various fields of science and engineering. (Detailed background on object classification and k-means applications can be found in the literature.) For instance, iterative k-means clustering is used to identify cancerous samples or to perform unsupervised learning tasks. The algorithm partitions a set of input data into k clusters, each of which is represented with a centroid. The original algorithm relies on the arithmetic mean to compute the centroids; therefore, the results are very sensitive to outliers. In response to this problem, more robust variants of the algorithm, such as aggregations and k-medoids, have been proposed and used in the past [10]. In particular, aggregations achieve better solution quality by setting the centroid of each cluster to its median. However, it requires excessive storage accesses to the data points and significantly limits the overall performance. Numerous techniques have been proposed in the literature to accelerate aggregations clustering, such as precise and approximate software solutions, field-programmable gate array accelerators using sorting networks, parallel probabilistic platforms, graphics processing unit accelerators , and application-specific hardware frameworks . Regrettably, the required data movement between the main storage and the processor cores limits the performance of these recent efforts even for moderately sized data sets. Moreover, with the growing interest in the future data intensive applications such as deep learning applications that rely on unsupervised classification the importance of high performance data clustering techniques is expected to increase.

The proposed hardware accelerator is evaluated on a k-means clustering library using breast cancer, indoor localization, and the data centric sets. Census data sets and two applications that use k-means clustering. We observe that the proposed accelerator achieves an average performance improvement of 2 times more than the present data set and an average data reduction of data twice over the soft-ware implementation of k-means on a processor system. Moreover, the results indicate an overall speedup of data centric with an data improvement of twice data centric over a processing-in-storage accelerator. When used for accelerating k-means clustering in gene expression analysis and classification based on term frequency–inverse document frequency, we observe that the proposed in situ accelerator can achieve speedups of data instances with respective data improvements of reduced data set compared with the processor baseline.

## 2. Objectives

An Data clustering is a computationally refers to partitioning a set of objects into meaningful groups (called clusters) with no predefined labels. The entities of a cluster are more similar to each other than to those in other clusters. K-means and its variants are the most prominent clustering algorithms that have been successfully used in numerous fields of science and engineering. The basic k-means operations are shown in Algorithm 1, where k centroids are used to represent the clusters. A centroid is either a representative member of the cluster (e.g., median) or an additional data point computed based on the similarities among all of the cluster members (e.g., the arithmetic mean). The former has been proven to find better clusters than the latter due to its resistance against outliers. Prior to partitioning, the k centroids are randomly initialized; then, the algorithm repeats two steps (Lines 3 and 4 in Algorithm 1) until convergence is reached. First, the clusters are formed by assigning data points to their closest centroid; second, the centroids are recomputed for each cluster.

Example Applications of Ddata clustering: Numerous applications of k-means clustering can be found in the literature. We review two representative examples on and text data mining.

a) data layer 1:

b) data layer 2:

Data layer1: Recently, clustering has seen wide use in medical research, such as cancer diagnosis and drug discovery. An accurate clustering algorithm often has a profound impact on the correctness of these applications. For example, Lu and Han have shown that data clustering algorithms can be used for more accurate cancer classifications based on the abundance of gene expression data rather than the traditional morphological and clinical-based methods. A gene that forms the basic unit of heredity is defined as part of a deoxyribonucleic acid transferred to an offspring by its parent. The process of transcribing a gene's sequence into ribonucleic acid is called gene expression that changes during biological phenomena, such as cell development. In the case of diseases such as cancer, the genes of normal body cells undergo multiple mutations to evolve cancerous cells. As shown in Fig. 1, this anomaly is now possible to be detected through data sets that require clustering of a large number of gene samples.

Data layer2: Clustering text documents is an important branch of text mining that refers to organizing paragraphs, sentences, and terms into meaningful clusters to improve information retrieval and document browsing [24]. Unlike the numerical data, text clustering requires preprocess-ing the documents to represent their features in the form of numerical vectors (see Fig. 2). These vectors are then used to group similar terms into the same clusters. A commonly used feature vector for text clustering is the TF, which represents the number of word occurrences in every document divided by the total number of words. Moreover, an IDF for every word

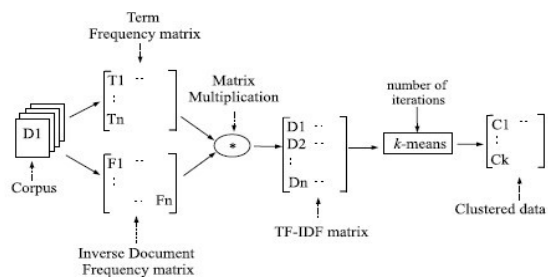

**Figure1. K-Means algorithm with frequency matric.**

Data clustering Using data layers with Filters Of non-polynomial and nonlinear digital components widely used in signal and image processing to filter out noise from input signals. In general, an data set is characterized by:

1) the number of data set (d) and

2) an index (N) that determines which input sig-nal to appear in the output.

The filter identifies the ith largest (or smallest) input signal to be sent to the output. In particular, a median filter can be realized if i = (N/2)

In principle, the median of a list can be computed using a sorting algorithm, which is complex and inefficient. Thereafter, numerous hardware and software implementations of bit-serial median filters have been proposed in the literature that relies on the majority function. The majority function defines a mapping from N binary inputs to a single output, such that the output is 0 when (N/2) or more inputs are 0; otherwise, it is 1.

| | A | B | C | D | E | F | G | H | I |
|---|---|---|---|---|---|---|---|---|---|
| 1 | State | Total Pop | Net Dome | Federal/C | Net Int. M | Period Bi | Period De | < 65 Pop. | > 65 Pop. |
| 2 | Alabama | 4464356 | -1.78 | -0.02 | 0.69 | 14.41 | 10.28 | 869.21 | 130.79 |
| 3 | Alaska | 634892 | -1.72 | -0.24 | 2.09 | 15.95 | 4.64 | 941.95 | 58.05 |
| 4 | Arizona | 5307331 | 14.25 | -0.03 | 4.29 | 15.88 | 7.77 | 869.54 | 130.46 |
| 5 | Arkansas | 2692090 | 0.36 | -0.01 | 1.07 | 14.35 | 10.51 | 861.06 | 138.94 |
| 6 | California | 34501130 | -2.01 | -0.04 | 7.88 | 15.37 | 6.72 | 894.03 | 105.97 |

**Table1 Data set**

Figure shows how to compute the median of five input numbers using the bit-serial algorithm. Every number is represented in a binary format within a single row. Starting from the most significant bit toward the instances of periods, the algorithm performs a vertical computation followed by a horizontal bit propagation, repeatedly.2 During the vertical computation, the majority vote among all of the bits within a selected data is computed. The result of the majority function is used: 1) to determine a bit of the final result to identify the minority bits within the selected data. In the next step, the minority bits are used to replace all of the bits on their right-hand side.

Implementation challenges: Theoretically, the bit-serial median algorithm is amenable to massively parallel implementations: selected bits from all of the inputs can be processed in parallel. In practice, however, this potential parallelism is significantly constrained due to the required excessive storage accesses to the input numbers per every iteration. This paper designs a storage-centric accelerator that performs the majority function computation and bit propagation steps in suitable storage arrays, thereby eliminating the unnecessary accesses to the inputs. As a result, the proposed platform will enable massively parallel execution of the bit-serial median algorithm.

## 3. Implementation

*D a t a c l u s t e r i n g w i t h d a t a l a y e r 1 :*

This section provides an overview of the proposed data set accelerator and explains the design principles for realizing data efficient data clustering within storage arrays. The key idea is to exploit the computational capabilities of the dataset elements in RRAM arrays to perform the necessary computation of the bit-serial median filter algorithm in storage cells. As a result, the proposed architecture eliminates unnecessary latency, bandwidth, and data overheads associated with streaming data out of the storage arrays during the clustering process. This novel capability will then unlock the unexploited massive parallelism in data clustering using bit-serial median filters.

Computing bit-serial median requires multiple steps, each of which involves activating the cells using bit lines and word lines. The assumed adjacent storage cells in a row. On every iteration, only one cell of each storage row will be processed. Based on P and I are initialized to determine if the cell should be included in computation. This is necessary to ensure that irrelevant data points are not included in computing the median value. To compute the majority vote of data $b_i$, sets C from datas $b_i$ and $b_{i-1}$ are connected to ground and set.

*D a t a a g g r e g a t i o n w i t h d a t a l a y e r 2 :*

A successive data set approximation mechanism is designed and employed to accomplish the analog bit counting for every data reset characteristics. To compute the majority vote of a data, is supplied to the compute set using a line driver. A two-level amplification mechanism consisting a current mirror and a current-mirror-based differential amplifier is employed to quantize the total current of the compute set.

Table 2 Data sending after reliability mechanism in layer2.

From given data we are Solving a large-scale data clustering problem requires computing the majority vote of a large number of data points stored in a single storage array, which becomes impractical due to significant sensing and reliability issues. Instead, data points are stored in multiple limited sized arrays, and only a fraction of the cells within each column is processed in every iteration cells of a array. Notice that multiple such operations are performed in parallel data arrays to achieve significant performance improvements. As shown in Fig. 9, a hierarchical merging mechanism is proposed to compute the majority vote of a large number of data points stored in numerous data arrays across the accelerator chip.5 An interconnection tree comprising reduction units is employed to merge the partial counts into a single value for computing the majority vote. The main purpose of the reduction tree is to merge the partial counts computed by the analog bit counters.

## 4. Process Investigations

Our experiments indicates that the data and delay overheads of this required preprocessing are negligible compared with the data and delay of transferring data to/from the accelerator and performing the actual clustering computations. Moreover, we observed that a 64-bit fixed point format for the evaluated applications and data sets achieves virtually the same results obtained with a double precision IEEE floating point format. Nevertheless, for sensitive applications, the proposed accelerator is flexible enough to compute the medians of wider bit representations by increasing the number of vertical majority vote computation and applying minimal changes to the control logic. Fig. 14 shows an example of floating point to fixed-point conversion. The input floating point data are scaled by a factor of 23 and then are converted to fixed-point data. We now apply an data set to compute the median of the fixed-point data.

We evaluate the accelerator for a k-means library with three real data sets belonging to different domains. First, data set profiles breast cancer samples containing nearly 6 data set sample. Second, an indoor localization data set with nearly 20 training records validated with more than 10 points, and the last data set contains nearly 1% sample of the total . census from]. We further assess the proposed hardware on two k-means-septic applications and text mining using summarizes the data sets used for the library and applications.

K-Mean data layer computations with segmented data set.

Steps as follows:

Specifically, the k-means algorithm shows as follows,

(1) Select $k$ items randomly as the initial clustering center $c_1, c_2, \ldots, c_k$;

(2) For the rest $(n-k)$ items, if $d_{ij}(x_i, c_j) < d_{im}(x_i, c_m)$, then $x_i \in C_i$;

(3) Calculate the centroid for each cluster $c_i^* = \frac{1}{n} \sum_{x \in C_i} x$

(4) If $c_i^* = c_i$ for all $i \in [1, k]$, algorithm terminates and $c_1^*, c_2^*, \ldots, c_k^*$ is the result. Otherwise let $c_i = c_i^k$ and GOTO (2)

K-Means is a well-known approach for unsupervised learning method to solve the clustering problem. It is iterative method starts with K initial cluster centers and each pattern is assigned to one of the pre-selected cluster centers based on its similarity. The iterative process converges when the cluster centers are very close in two successive iterations. This method takes O (tkn) time complexity, where n is the number of objects, k is the number of clusters, and t is how many iterations it takes to converge. In case of large values of n the method takes long time to converge. Hence many techniques were proposed to reduce the number of distance computations and thereby reduce the running time. Few attempts were made to improve the conventional k-means . Some prototype based hybrid clustering methods have been proposed to speedup the k-means method to work with Big Data. This paper presents a two level prototype based hybrid approach to speedup the k-means method for Big Data.

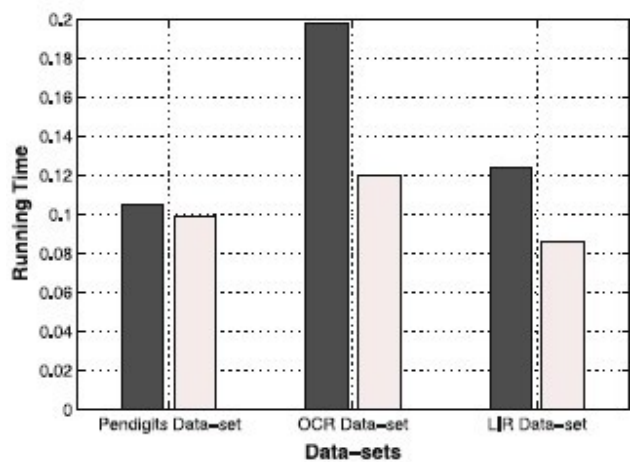

**Figure 2. computing ratios of each instances**

(1) determine the cluster count $[k_{min}, k_{max}]$
(2) while $k_{min} < k < k_{max}$
(2.1) call the traditional k-means algorithm
(2.2) get the BMP for each sample
(2.3) calculate the $avgBMP(k)$
(3) get the optimal cluster count $k_{opt}$
(4) output the result

Classification: Classification, as the name suggests, is the process of placing or assigning the categorical variables into predefined classes. An algorithm needs to be selected to place these data in categories. The decision rules are based on training data and then used to locate these data in pre-determined groups. The rules are further validated by the validation dataset.

B. Clustering: In clustering, the data items are clustered according to their logical relationships or natural groupings and a structure as a whole is generated. There are no pre-defined groups, thus, clustering comes in the group of undirected Data Mining techniques. Each cluster is collection of homogeneous elements, which may be exclusive to that group, but are similar to each other. K-means clustering is a simple clustering method that has been used in similar research.

C. Prediction: Predicting the future can be possible if we have enough data to work with. Especially, using data mining. Data can be mined to foretell trends, patterns and behaviors. To form a foundation for the prediction model, the previously generated decision rules which are obtained using classification or clustering are used. This is the main idea behind prediction. Therefore, all previously discussed techniques have, to some extent, the capability of prediction.

D. Sequence Discovery: This technique determines sequential patterns. Datasets can often contain many sequential patterns which are usually in the form of K means clustering is one of the basic clustering algorithms in the machine learning domain. The inference of this algorithm is based on the value of „k" which is the number of clusters that can be found in an n-dimensional dataset. Usually the value of „k" is assumed or known a-priori. In k-means algorithm, since it is considered that there is „k" number of clusters; we consider that there are „k" number of cluster means (cluster centers), where the cluster mean is the average of all the data-points falling under each cluster. The end result of the k-means clustering algorithm is that each data point in the data-set is grouped into „k" clusters around the „k" cluster means. If the data points are tightly surrounding the cluster means, then it is considered as a highly cohesive and good cluster, else it is not. Hence the cluster compactness forms the metric of quality of the k-means algorithm.

k-means is one of the simplest unsupervised learning algorithms that solve the well known clustering problem. The procedure follows a simple and easy way to classify a given data set through a certain number of clusters (assume k clusters) fixed apriority. The main idea is to define k centers, one for each cluster. These centers should be placed in a cunning way because of different location causes different result. So, the better choice is to place them as much as possible far away from each other. The next step is to take each point belonging to a given data set and associate it to the nearest center. When no point is pending, the first step is completed and an early group age is done. At this point we need to re-calculate k new centroids as barycenter of the clusters resulting from the previous step. After we have these k new centroids, a new binding has to be done between the same data set points and the nearest new center. A loop has been generated. As a result of this loop we may notice that the k centers change their location step by step until no more changes are done or in other words centers do not move any more. Finally, this algorithm aims at minimizing an objective function know as squared error function given by:

Algorithmic steps for k-means clustering

Let $X = \{x1, x2, x3, \ldots, xn\}$ be the set of data points and $V = \{v1, v2, \ldots, vc\}$ be the set of centers.

1) Randomly select 'c' cluster centers.

2) Calculate the distance between each data point and cluster centers.

3) Assign the data point to the cluster center whose distance from the cluster center is minimum of all the cluster centers..

4) Recalculate the new cluster center using:

$$v_i = (1/c_i) \sum_{j=1}^{c_i} x_i$$

where, '$c_i$' represents the number of data points in ith cluster.

5) Recalculate the distance between each data point and new obtained cluster centers.

6) If no data point was reassigned then stop, otherwise repeat from step 3).

Advantages

1) Fast, robust and easier to understand.

2) Relatively efficient: O(tknd), where n is # objects, k is # clusters, d is # dimension of each object, and t is # iterations. Normally, k, t, d << n.

3) Gives best result when data set are distinct or well separated from each other.

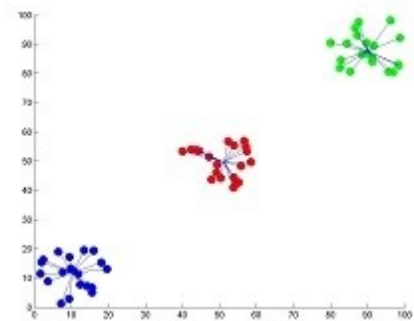

Figure3. Showing the result of k-means for 'N' = 60 and 'c' = 3

Note: For more detailed figure for k-means algorithm please refer to k-means figure sub page.

Disadvantages

1) The learning algorithm requires apriori specification of the number of cluster centers.

2) The use of Exclusive Assignment - If there are two highly overlapping data then k-means will not be able to resolve that there are two clusters.

3) The learning algorithm is not invariant to non-linear transformations i.e. with different representation of data

we get different results (data represented in form of cartesian co-ordinates and polar co-ordinates will give different results).

4) Euclidean distance measures can unequally weight underlying factors.

5) The learning algorithm provides the local optima of the squared error function.

6) Randomly choosing of the cluster center cannot lead us to the fruitful result. Pl. refer Fig.

7) Applicable only when mean is defined i.e. fails for categorical data.

8) Unable to handle noisy data and outliers.

9) Algorithm fails for non-linear data set.

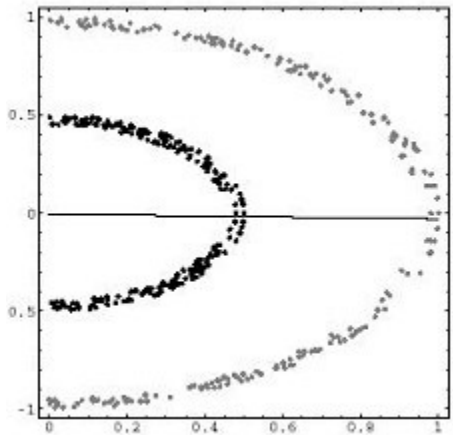

**Figure4. Showing the non-linear data set where k-means algorithm fails**

Nature-inspired optimization algorithms require certain functional parameters to be initiated with values to run. The function parameters are defined as follows. They allow the users to set with user-specific values for customizing the operations of the algorithms. Some of the parameters are common across different bio inspired optimization algorithms. In this paper, we have four hybrids, which resulted from combining four bio inspired optimization algorithms into K-means.

The goal of this clustering algorithm is to search for the best center to minimize the distance between the center of the cluster and its points. The full length of dataset is used for training—in clustering, building clusters are referred to until perfection is attained using the full set of data. Performance of the clustering is evaluated in terms of cluster integrity which is reflected by the intra- and inter similarities of data points within and across different clusters, the average sum of CPU time consumption per iteration during the clustering operation, and the number of loops taken for all the clusters to get converged. The criterion for convergence which decides when the looping of evolution stops is the fraction of the minimum distance between the initial cluster centers.

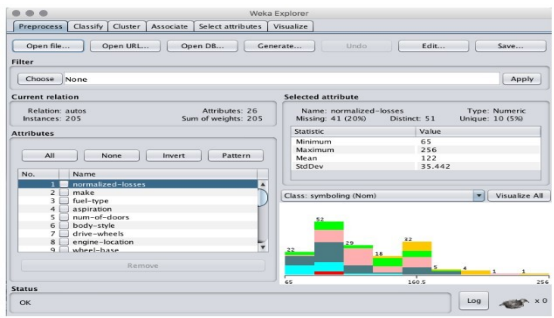

**Figure5. Explorer Interface**

In practical scientific and industrial applications, the quality and accuracy of image segmentation are very important which depend on the underlying data clustering algorithms. A common choice of unsupervised clustering algorithm is K-means in image segmentation based on color. The regions of the image depending on the color features are grouped into a certain set of segments by measuring the intercluster distance and intracluster distance between each image pixel and the centroid within the cluster. The clustering process is exactly the same as that used in the previous experiment on UCI datasets, except that the images under test in this experiment are larger in amount.

K-Means clustering algorithms, a classical class of partition-based algorithms used for merging similar data into clusters, are known to have the limitation of getting stuck in local optima. As a matter of intellectual curiosity in computer science, how best to cluster data such that the integrity of the clusters is maximized, has always been a challenging research question. The ideal solution is to find an optimal clustering arrangement which is globally best—so that no other possible combinations of data clustering exist that is better than the global one. One way of achieving this is to try all the possible combinations by brute-force which could be computational intractable. Alternatively, nature-inspired optimization algorithms, which recently rise as a popular research topic, are extended to work with K-means in guiding the convergence of disparate data points and to steer them towards global optima, stochastically instead of deterministically. These two research directions of metaheuristic optimization and data mining do fit like hand and glove. Constrained by the inherent limitation of K-means design and the merits of nature-inspired optimization algorithms, it is feasible to combine them letting them complement and function together.

When examining serial K-means algorithm, it can be observed that the algorithm deals with all objects in dataset serially which very time consuming especially for large databases. When huge datasets are in account, serial K-means algorithm either lacks in performance or crashes because of the larger dataset than the amount of memory of a single machine.

Attributes taken for clustering :

fixed acidity;"volatile acidity";"citric acid";"residual sugar";

"chlorides";"free sulfur dioxide";"total sulfur dioxide";

"density";"pH";"sulphates";"alcohol";"quality"

Instances taken for clustering:

7.4;0.7;0;1.9;0.076;11;34;0.9978;3.51;0.56;9.4;5

7.8;0.88;0;2.6;0.098;25;67;0.9968;3.2;0.68;9.8;5

7.8;0.76;0.04;2.3;0.092;15;54;0.997;3.26;0.65;9.8;5

11.2;0.28;0.56;1.9;0.075;17;60;0.998;3.16;0.58;9.8;6

7.4;0.7;0;1.9;0.076;11;34;0.9978;3.51;0.56;9.4;5

7.4;0.66;0;1.8;0.075;13;40;0.9978;3.51;0.56;9.4;5

7.9;0.6;0.06;1.6;0.069;15;59;0.9964;3.3;0.46;9.4;5

7.3;0.65;0;1.2;0.065;15;21;0.9946;3.39;0.47;10;7

7.8;0.58;0.02;2;0.073;9;18;0.9968;3.36;0.57;9.5;7

7.5;0.5;0.36;6.1;0.071;17;102;0.9978;3.35;0.8;10.5;5

6.7;0.58;0.08;1.8;0.097;15;65;0.9959;3.28;0.54;9.2;5

7.5;0.5;0.36;6.1;0.071;17;102;0.9978;3.35;0.8;10.5;5

5.6;0.615;0;1.6;0.089;16;59;0.9943;3.58;0.52;9.9;5

7.8;0.61;0.29;1.6;0.114;9;29;0.9974;3.26;1.56;9.1;5

8.9;0.62;0.18;3.8;0.176;52;145;0.9986;3.16;0.88;9.2;5

8.9;0.62;0.19;3.9;0.17;51;148;0.9986;3.17;0.93;9.2;5

8.5;0.28;0.56;1.8;0.092;35;103;0.9969;3.3;0.75;10.5;7

8.1;0.56;0.28;1.7;0.368;16;56;0.9968;3.11;1.28;9.3;5

7.4;0.59;0.08;4.4;0.086;6;29;0.9974;3.38;0.5;9;4

Observations as follows:

Success of data mining is strongly related with data warehousing functions of companies. Data mining uses pre-processed data which are supplied by data warehouses. Companies' data warehousing departments develop functions which collect valuable data from business activities continuously. Collected and cleansed data are then stored in data warehouses in order to be used in statistical works. Data mining is the one of the most important ones of those statistical works which uses the data stored in data warehouses.

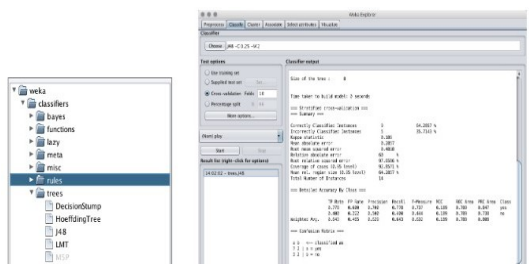

**Figure6.Data set in classifier list.**

When examining results of different runs with different sets of random initial points, it has been observed that selection of initial points also affects the convergence time of K-means algorithm. In one example, execution time of the algorithm has become the half of the previous run time by the change of random initial points. This shows that, a technique for selecting better initial points than random ones may be developed and used in connection with the parallel algorithm as a future work in order to make K-means algorithm even faster.

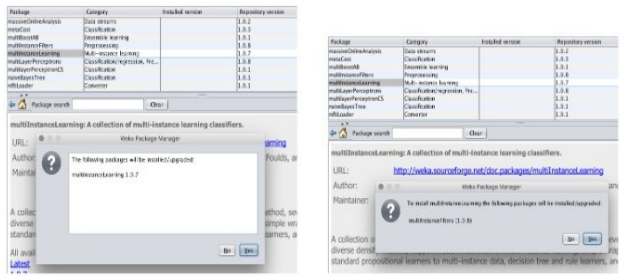

**Figure7. Illustration of data set loading.**

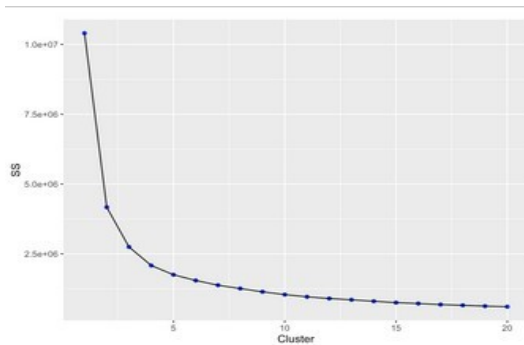

**Figure8. K-mean clustering on data set.**

|  | fixed_acidity | volatile_acidity | citric_acid | residual_sugar |
|---|---|---|---|---|
| nbr.val | 4898.0000 | 4898.00000 | 4898.00000 | 4898.0000 |
| nbr.null | 0.0000 | 0.00000 | 19.00000 | 0.0000 |
| nbr.na | 0.0000 | 0.00000 | 0.00000 | 0.0000 |
| min | 3.8000 | 0.08000 | 0.00000 | 0.6000 |
| max | 14.2000 | 1.10000 | 1.66000 | 65.8000 |
| range | 10.4000 | 1.02000 | 1.66000 | 65.2000 |
| sum | 33574.7500 | 1362.52500 | 1636.87000 | 31305.1500 |
| median | 6.8000 | 0.26000 | 0.32000 | 5.2000 |
| mean | 6.8548 | 0.27824 | 0.33419 | 6.3914 |
| SE.mean | 0.0121 | 0.00144 | 0.00173 | 0.0725 |
| CI.mean.0.95 | 0.0236 | 0.00282 | 0.00339 | 0.1421 |
| var | 0.7121 | 0.01016 | 0.01465 | 25.7255 |
| std.dev | 0.8439 | 0.10079 | 0.12102 | 5.0721 |
| coef.var | 0.1231 | 0.36226 | 0.36213 | 0.7936 |

|  | chlorides | free_sulfur_dioxide | total_sulfur_dioxide |
|---|---|---|---|
| nbr.val | 4898.000000 | 4898.000 | 4898.000 |
| nbr.null | 0.000000 | 0.000 | 0.000 |
| nbr.na | 0.000000 | 0.000 | 0.000 |
| min | 0.009000 | 2.000 | 9.000 |
| max | 0.346000 | 289.000 | 440.000 |
| range | 0.337000 | 287.000 | 431.000 |
| sum | 224.193000 | 172939.000 | 677690.500 |
| median | 0.043000 | 34.000 | 134.000 |
| mean | 0.045772 | 35.308 | 138.361 |
| SE.mean | 0.000312 | 0.243 | 0.607 |
| CI.mean.0.95 | 0.000612 | 0.476 | 1.190 |
| var | 0.000477 | 289.243 | 1806.085 |
| std.dev | 0.021848 | 17.007 | 42.498 |
| coef.var | 0.477318 | 0.482 | 0.307 |

## 5. Comparisons of Data Layers with Data set analysis and conclusions

According to the algorithm we firstly select k objects as initial cluster centers, then calculate the distance between each cluster center and each object and assign it to the nearest cluster, update the averages of all clusters, repeat this process. To summarize, there are different categories of clustering techniques including partitioning ,hierarchical, density-based and grid-based clustering. The K-means clustering algorithm is a clustering technique that falls into the category of partitioning.

The algorithm finds a partition in which data points within a cluster are close to each other and data points in different clusters are far from each other as measured by similarity. As in other optimization problems with a random component, the results of k-means clustering are initialization dependent. This is usually dealt with by running the algorithms several times, each with a different initialization. The best solution from the multiple runs is then taken as the final solution.

| Dataset | Recognition Rate (%) Number of cluster Formed | | | | |
|---|---|---|---|---|---|
|  | 3 | 5 | 10 | 14 | 16 |
| Iris | 88.66 | 97.33 | 98 | 99.33 | 99.33 |
| Wine | 70.29 | 73.42 | 75.73 | 77.05 | 78.31 |
| Vowel | - | - | 69.04 | 75.02 | 72.90 |
| Ionosphere | 79.30 | 80.22 | 83.61 | 87.15 | 88.55 |
| Crude oil | 60.52 | 73.09 | 89.07 | 89.95 | 93.46 |

**Table 3 clustering dataset**

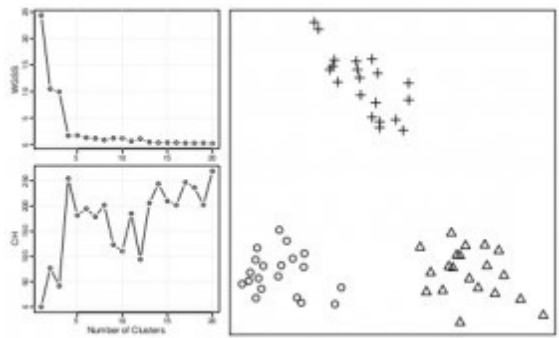

**Figure10. Comparisons of each data**

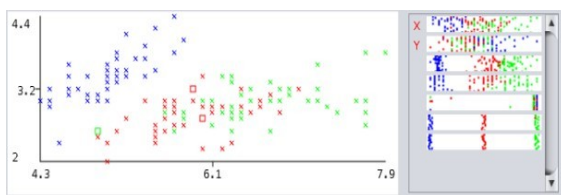

**Figure11. ploting at data layer1.**

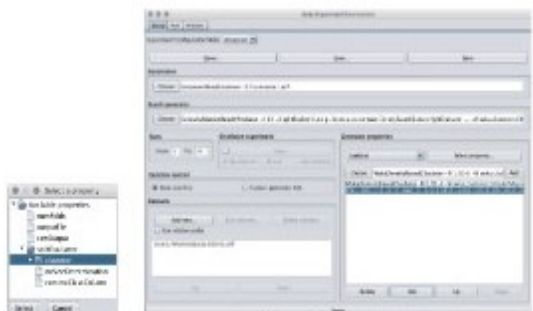

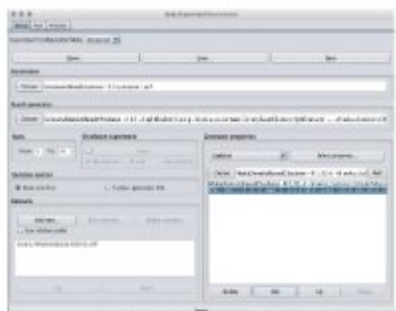

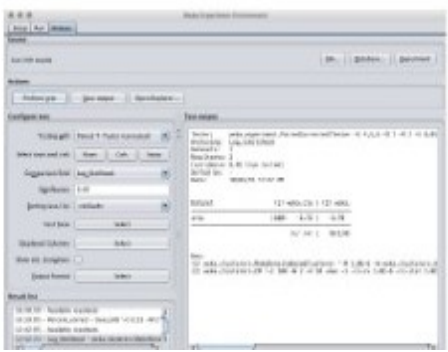

**Figure12. Experimental view of each clustering**

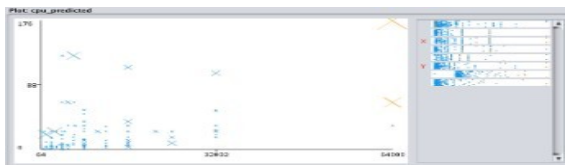

**Figure13.ploting at data layer2**

## 6. Command line interface of data layers

When cross-validation is performed (the default if a test file is not provided), the data is randomly shuffled first. To repeat the cross-validation several times, each time reshuffling the data in a different way, set the random number seed with –s(default value 1). With a large dataset you may want to reduce the number of folds for the cross-validation from the default value of 10 using –x. If performance on the training data alone is required,–no-cvan be used to suppress cross-validation;–v suppresses output of performance on the training data. As an alternative to cross-validation, a train-test split of the data specified with the –top portion can be performed by supplying a percentage to use as the new training set with–split-percentage (there maintaining data is used as the test set). Randomization of the data can be suppressed when performing a train-test split by specifying–preserve-order

```
2 2    % Number of rows and columns in the matrix
0 10   % If true class yes and prediction no, penalty is 10.
1 0    % If true class no and prediction yes, penalty is 1.
```

The first line gives the number of rows and columns, that is, the number of class values. Then comes the matrix of penalties. Comments introduced by % can be appended to the end of any line. It is also possible to save and load models. If you provide the name of an output file using WEKA saves the classifier generated from the training data. To evaluate the same classifier on a new batch of test data, you load it back using –listed of rebuilding it. If the classifier can be updated incrementally, you can provide both a training file and an input file, and WEKA will load the classifier and update it with the given training instances. If you wish only to assess the performance of a learning scheme, use –o to suppress output of the model. Use –k to compute information-theoretic measures from the probabilities derived by a learning scheme.

| Option | Function |
|---|---|
| -U | Use unpruned tree |
| -O | Do not collapse tree |
| -C <pruning confidence> | Specify confidence threshold for pruning |
| -M <number of instances> | Specify minimum number of instances in any leaf |
| -R | Use reduced-error pruning |
| -N <number of folds> | Specify number of folds for reduced-error pruning; one fold is used as pruning set |
| -B | Use binary splits only |
| -S | Don't perform subtree raising |
| -L | Retain instance information |
| -A | Smooth the probability estimates using Laplace smoothing |
| -J | Prevent the use of the MDL correction for the information gain of numeric splits |
| -Q | Seed for shuffling the data |
| -doNotMakeSplitPointActualValue | Do not find a split point that is an actual value in the training set |

**Table 4 Schemes for specific options**

Internally, an Instance stores all attribute values as double-precision floating-point numbers regardless of the type of the corresponding attribute. In the case of nominal and string attributes this is done by storing the index of the corresponding attribute value in the definition of the attribute. For example, the first value of a nominal attribute is represented by 0.0, the second by 1.0, and so on.

Data source Link: https://archive.ics.uci.edu/ml/machine-learning-databases/wine-quality/

The proposed method can be viewed as an algorithm for initializing an iterative partitioning algorithm like k-means, which is the focus of this paper. k-means is on e of the simplest and most popular approaches for solving clustering problems despite its severe sensitivity to initialization. While k-means appears as a final step in the proposed algorithm, other partitioning algorithms could be used.

In k-means clustering k points are randomly chosen as the initial cluster centers. The choice of the number k of clusters is usually based on some heuristic. Each data point is assigned to the group that has the closest center. The cluster centers of mass are then recomputed. The assignment and re-computation steps are iterated until the intra - cluster variance converges to a minimum. K-means algorithm is to place every cluster centroid at the modes of the joint probability density of the data. Motivated by this idea, Bradley and Fayyad [28] propose a refinement of initial condition ns of k-means near the modes of the estimated distribution, by the use of a recursive procedure executing k-means on small random sub-samples of the data. Global k-means was introduced by Likas et al. and is also motivated by the same idea that modes play an important role. Their method is based on a recursive partitioning of the data space into disjoint subspaces by using k−d trees. They then define a cutting hyper plane as a linear space perpendicular to the highest variance axis. This is similar to the data partitioning that occurs in adaptive vector quantization and regression trees.

## 7. Conclusion

In K-means and aggregations clustering are widely used techniques for data clustering in scientific research and engineering disciplines. Most of the available solutions suffer in performance and data due to excessive data movement involved throughout the clustering process. The proposed data clustering accelerator successfully addresses these concerns by implementing bit-serial 2 data layer scheme for median calculation and performing in the data occurrence computations within the data cells, thereby eliminating unnecessary data movement between the core and the main storage. Based on our simulation results, the proposed accelerator achieves significantly better data and performance improvements as compared with processor and accelerators. In conclusion, the proposed hardware accelerator significantly improves the performance and data for clustering applications involving processing of very large data sets.

**References**

[1] Cheng-Tao Chu, Sang Kyun Kim, Yi-An Lin, YuanYuan Yu, GaryBradski, Andrew Y. Ng and Kunle Olukotun, "Map-Reduce for MachineLearning on Multicore", Proceeding NIPS'06 Proceedings of the 19thInternational Conference on Neural Information Processing Systems, pp.281-288, 2006.

[2] C.L. Philip Chen and Chun-Yang Zhang, "Data-intensive applications,challenges, techniques and technologies: A survey on Big Data",Information Sciences, vol.275, pp.314-347, August 2014.

[3] Weizhong Zhao, Huifang Ma and Qing He, "Parallel K-Means Clustering Based on MapReduce", Lecture Notes in Computer Science,vol.5931, pp.674-679, 2009.

[4] Muhammad Bilal, Lukumon O. Oyedele, Olugbenga O. Akinade,Saheed O. Ajayi, Sururah A. Bello," Big data architecture for construction waste analytics (CWA): A conceptual framework",Journal of Building Engineering, vol. 6, pp. 144-156, June 2016.

[5] Yichuan Wang and Nick Hajli, "Exploring the path to big data analyticssuccess in healthcare", Journal of Business Research, vol. 70, pp. 287–299, January 2017.

[6] Zhenlong Li, Qunying Huang, Gregory J. Carbone, Fei Hu,"A highperformance query analytical framework for supporting data-intensiveclimate studies",Computers, Environment and Urban Systems, vol. 62,pp.210-221, March 2017.

[7] Jui-Sheng Chou, Ngoc-Tri Ngo,"Smart grid data analytics frameworkfor increasing energy savings in residential buildings", Renewable andSustainable Energy Reviews,volume 66, pp. 499–516,December 2016.

[8] Ali Dag, Asil Oztekin, Ahmet Yucel, Serkan Bulur, Fadel M. Megahed,"Predicting heart transplantation outcomes through data analytics", Decision Support Systems, vol. 94, pp. 42-52, February 2017.

[9] D. Xia, H. Li, B. Wang, Y. Li and Z. Zhang, "A Map Reduce-BasedNearest Neighbor Approach for Big-Data-Driven Traffic FlowPrediction," IEEE Access, vol. 4, no. , pp. 2920-2934, 2016.

[10] A. Paul, A. Ahmad, M. M. Rathore and S. Jabbar, "Smartbuddy:defining human behaviors using big data analytics in social internet ofthings," IEEE Wireless Communications, vol. 23, no. 5, pp. 68-74,October 2016.